\documentclass[twocolumn,aps,prl,amsmath,longbibliography]{revtex4-1}

\usepackage{graphicx}
\usepackage{dcolumn}
\usepackage{color}
\usepackage{hyperref}
\hypersetup{colorlinks=true,citecolor=red}

\newcommand{\EA}[1]{{\color{black}#1}}

\newcommand{\EAQ}[1]{{\color{black}#1}}

\begin{document}
	
	\graphicspath{{./figs/}}
	\title{State-of-the-art and prospects for intense red radiation from core-shell InGaN/GaN nanorods}
	\author{E. A. Evropeitsev$^1$}
	\email{evropeitsev@beam.ioffe.ru}
	\author{D. R. Kazanov$^1$}
	\author{Y. Robin$^{2}$}
	\author{A. N. Smirnov$^1$}
	\author{I. A. Eliseyev$^1$}
	\author{V.~Yu.~Davydov$^1$}
	\author{A. A. Toropov$^1$}
	\author{S. Nitta$^{2}$}
	\author{T. V. Shubina$^1$}
	\author{H. Amano$^2$}
	
	\affiliation{$^1$Ioffe Institute, 26 Politekhnicheskaya, St Petersburg 194021, Russia}
	\affiliation{$^2$Institute of Materials and Systems for Sustainability (IMaSS), Nagoya University, Nagoya, Japan}

	\date{\today}
	
\begin{abstract}
	Core-shell nanorods (NRs) with InGaN/GaN quantum wells (QWs) are promising for monolithic white light-emitting diodes and multicolor displays. Such applications, however, are still a challenge because intensity of red band is too weak as compared with blue and green ones. 
	To clarify the problem, we have performed power and temperature dependent, as well as time-resolved measurements of photoluminescence (PL) in NRs of different In content and diameter. These studies have shown that the dominant PL bands originate from nonpolar and semipolar QWs, while a broad yellow-red band arises mostly from defects in the GaN core. Intensity of red emission from the polar QWs at the NR tip is fatally small. Our calculation of electromagnetic field distribution inside the NRs shows a low density of photon states in the tip that suppresses the red radiation.  We suggest a design of hybrid NRs, in which polar QWs, located inside the GaN core, are pumped by UV-blue radiation of nonpolar QWs. Possibilities of radiative recombination rate enhancement by means of the Purcell effect are discussed.

\end{abstract}

	\pacs{Valid PACS appear here}
	
	\keywords{core-shell, nanorod, InGaN}
	\maketitle
	
	\section{Introduction}
	Core-shell nanorods (NRs), which comprise InGaN quantum wells (QWs) in a shell deposited over a GaN core, are among the most promising objects for modern nanophotonics \cite{Li2012}. It is assumed that monolithic NR-based devices can be used in multi-color displays and solid-state lighting \cite{Hong2011, Kim2014} 
	instead of commonly used phosphor coated light emission diodes (LEDs) \cite{Nakamura1997}.
	The InGaN/GaN QWs are a good basis for that, because they can emit light in any part of the visible range depending on the thickness and composition \cite{Nakamura1995,Ivanov2006}. In the core-shell NRs, the variation of these parameters is realized spontaneously as a result of different growth rate and In incorporation on the different crystallographic planes (Figure 1 (a)). 

	\begin{figure}[t] 
		\includegraphics[width=.99\columnwidth]{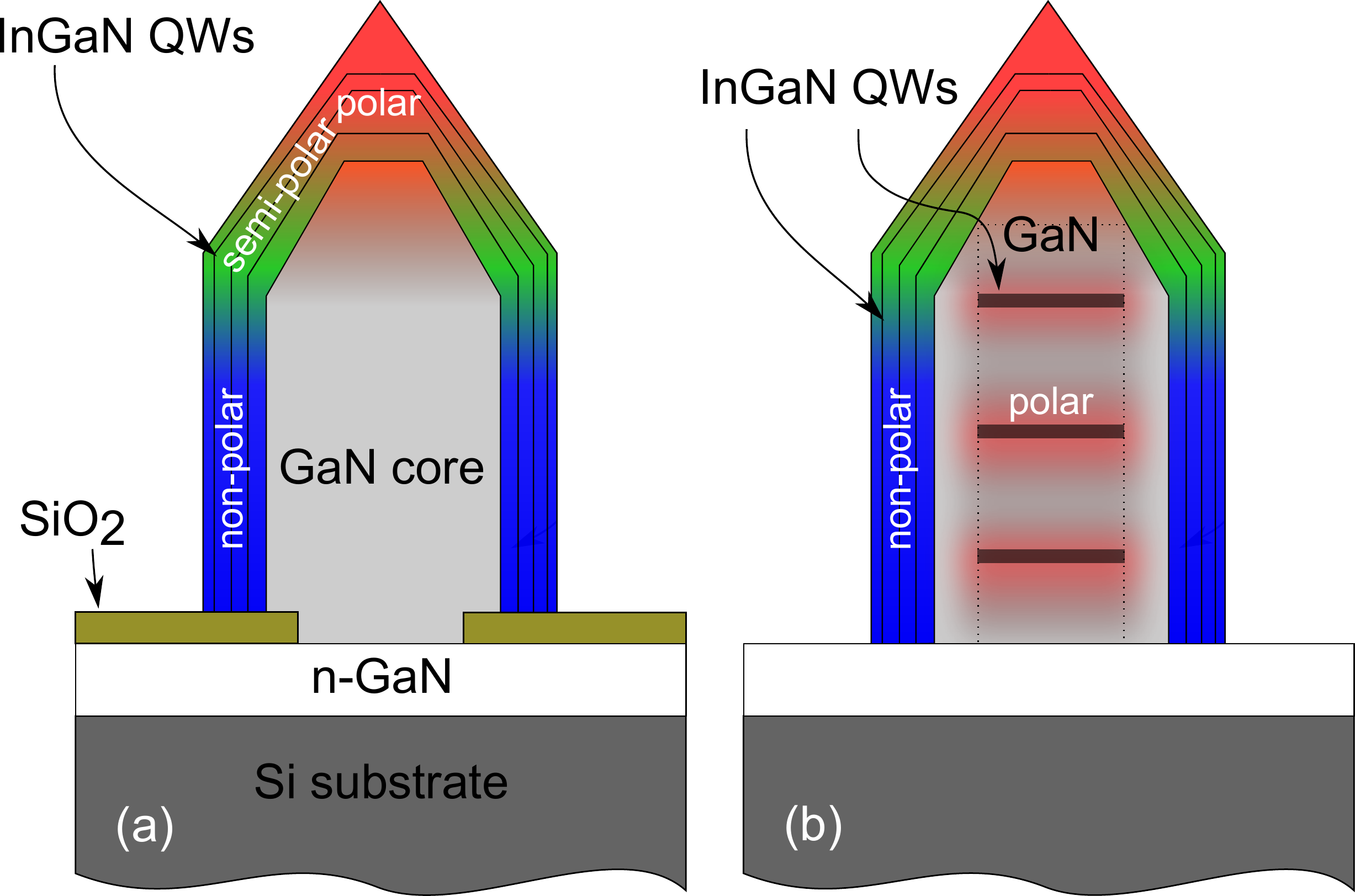}
		\caption{ Sketch of core-shell GaN/InGaN nanorod (a) and a proposed design of a hybrid NR structure (b).}\label{fig:Design}
	\end{figure}

	Actually, there are many obstacles to achieve these aims. To get white light, for instance, one should combine the light of two (Y, B) or three (R, G, B) wavelengths with comparable intensities. However, the internal quantum efficiency (IQE) of InGaN/GaN QWs is usually low in the yellow-red spectral range \cite{Mukai1999}. The InGaN radiation gets into this range when the well width is wide and In content $x\simeq$0.4-0.6. When the In content is high, the critical thickness is small and the phase separation occurs, inevitably deteriorating the QW quality \cite{Ho1996}. In addition, the built-in electric field across the polar or semipolar QWs spatially separates electrons and holes decreasing the IQE \cite{Kioupakis2012}. The overlap of electron-hole wave functions cannot exceed 1\% in polar InGaN/GaN QWs emitting at wavelengths longer than $\sim$560-600 nm \cite{Pristovsek2013}. The quantum-confined Stark effect (QCSE) is stronger in the wide InGaN/GaN QWs with high In content due to the increasing strain and concomitant piezoelectric polarization.  However, a positive feature of the QCSE is the shift of the emission towards the longer wavelengths that sometimes is crucial for obtaining red emission.
	
	Over the past two decades, various approaches to the manufacture of monolithic InGaN/GaN white LEDs have been reported. In several papers, the blue light, emitted from electrically-pumped active region, was partially converted into yellow-red photoluminescence (PL) by the QWs of different widths and compositions. More importantly, these QWs, situated outside the active region, were pumped only optically \cite{Damilano2001, Damilano2008, Damilano2010, Kowsz2015, Kowsz2016}. This approach makes it possible to compensate for the small efficiency of red radiation by increasing the number of the re-emitting QWs.  
	The alternative approach implies that the active region includes InGaN inserts -- quantum wells, disks, or dots of various width/composition/polarity designed to cover the entire visible range together. This approach dominates up to now \cite{Yamada2002, Lee2007, Wang2007, Lee2008, Lin2010, Guo2011, Wu2014, Lee2017, Kim2014, Lee2015}.  It is worth mention that the increase in the number of QWs when all of them are pumped electrically does not mean a proportional increase in electroluminescence (EL) intensity because of the limitation of carrier injection and transport \cite{David2008,Laubsch2010}.

	The state-of-the art core-shell NRs are designed in accordance with the last approach. Polar, semipolar, and nonpolar QWs, formed simultaneously, can emit in the red, green, and UV-blue regions, respectively \cite{Kim2014,Robin2018a}. The advantages of such three-dimensional (3D) structures over planar ones are numerous: low dislocation density, use of nonpolar planes, large effective surface area, and high light extraction efficiency \cite{Li2012}. The polar QWs at the top of NRs are usually wider and have higher In content relatively to semi- and nonpolar QW. Together with the strong QCSE in the polar QWs, this leads to the radiation at longer wavelengths \cite{Hong2011, Kusch2018, Robin2018b}. 
	Systematically, 
	 a weak red EL appears at low applied voltages, while blue/green radiation dominates with their increase \cite{Hong2011,Robin2018a}. Generally, the red radiation can be enhanced by using truncated pyramids \cite{Lee2015} or by increasing the NR diameter \cite{Robin2018b}. In this case, however, important advantages of the NRs such as dislocation filtration and large effective area may be lost and will cause the IQE drop \cite{Sekiguchi2019}.
	
	In this article, we report on studies of core-shell NRs with and without InGaN/GaN QWs. Temperature-dependent time-resolved PL (TRPL) spectroscopy and power-dependent PL allows us to attribute the two dominant PL peaks to nonpolar and semipolar QWs. We show how these peaks shift from UV to green spectral regions depending on the nominal QW composition, and demonstrate their relatively high IQE. In contrast, a broad yellow-red emission band occurs mainly due to defects in the GaN core.  We present a calculation of electromagnetic field distribution in NRs, which shows a negligible density of photon states at NR tips. This can lead to inhibition of red radiation, which was initially weak due to the small area of the polar QW and QCSE. To overcome these problems we offer a hybrid NR design (see Figure 1 (b)) in which the positive Purcell effect can be achieved. This implies the location of polar QWs in the core for \EAQ{efficient conversion} of UV-blue light from nonpolar QWs to red radiation.

	\section{Samples and methods}

	\begin{figure}[t] 
		\includegraphics[width=.99\columnwidth]{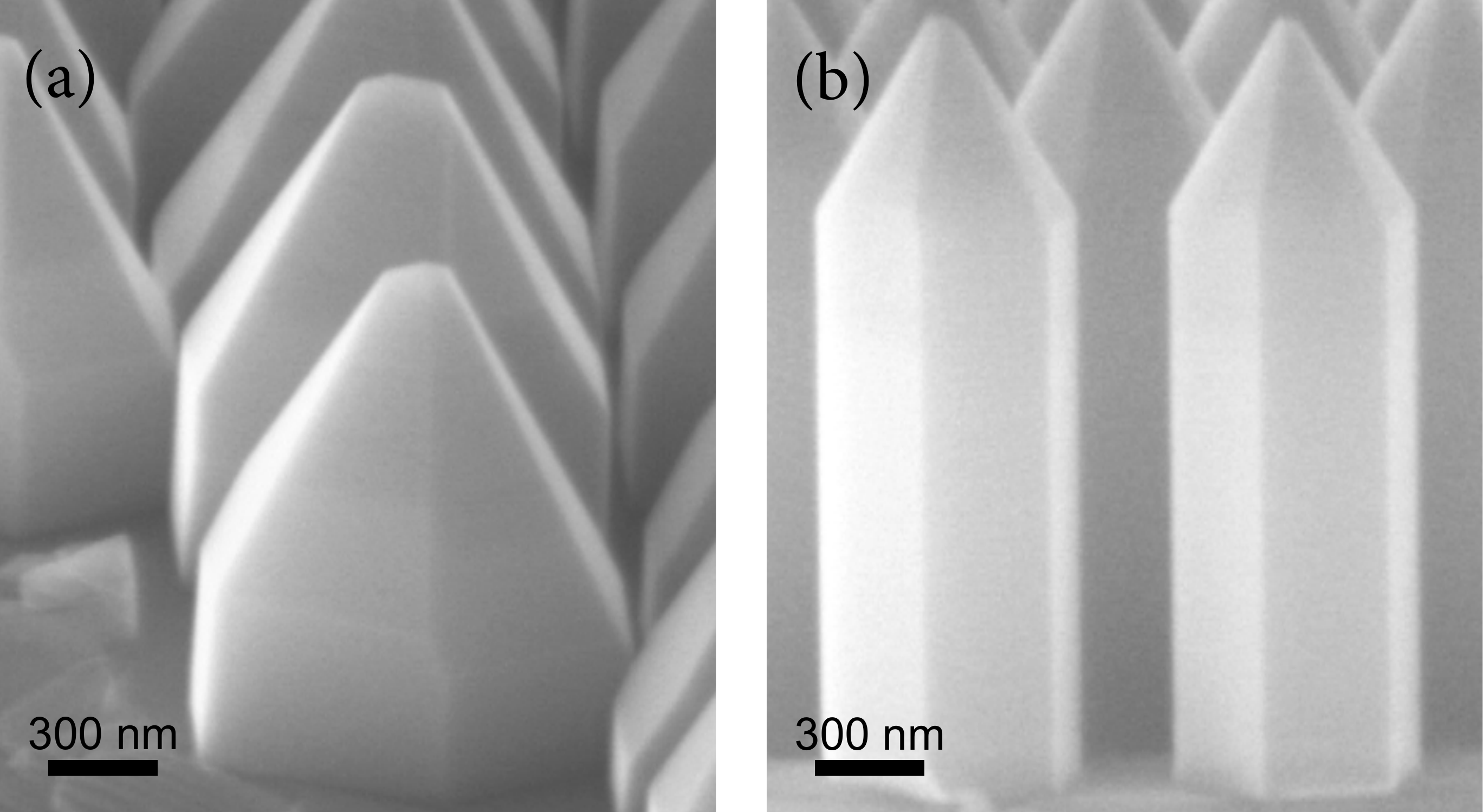}
		\caption{ Scanning electron microscopy images of two NR samples: A1 (a) and A2 (b).
		}\label{fig:SEM}
	\end{figure}

	In our studies, we have used the samples with arrays of NRs  grown by metal-organic vapor phase epitaxy (MOVPE). The growth procedure was as follows. At first, 750 nm i-GaN and 750 nm n-type GaN buffer layers were grown on a sapphire substrate. Then, a 30 nm thick $\rm{SiO_{2}}$ mask was fabricated by plasma-enhanced chemical vapor deposition (PECVD) for selective area epitaxy (SAE). An array of holes with two diameters (190 or 460 nm) was created in the $\rm{SiO_{2}}$ layer by a combination of nanoimprint lithography, photolithography, and dry etching techniques. Afterwards, the templates were loaded in the MOVPE growth chamber and the GaN/InGaN regrowth was carried out in the following sequence: the n-GaN core, three InGaN/GaN QWs and capping p-GaN outer shells. The growth conditions were similar to those described in \cite{Robin2018b}. The planar template (without a mask) was co-loaded to grow a reference polar QWs in the same conditions. The thickness of the wells and barriers in the planar samples were specified by X-ray diffraction as 3-3.5 nm and 15-16~nm, respectively. During the growth of the QWs only the temperature {$\rm {T_{gr}}$ was varied to change the wavelength of QW luminescence. The temperature was set to 800, 760 and 720$^{\circ}$C to provide nominal In composition in QWs in the planar samples A, B, and C as 12, 19 and 26 \%, respectively (see Table~\ref{table1}). Later, we will refer to the NR samples, grown with 460 nm holes in the mask, as A1, B1, and C1, and with 190 nm holes in the mask as A2, B2, and C2. In addition, the reference NR sample without any InGaN QWs (only core) was grown. 

%

\begin{table}[t]
	\begin{center}
		\renewcommand{\arraystretch}{1.2} 
		\centering
		\caption{Series of planar and nanorod samples.}
		\label{table1}
		\begin{tabular}{@{}m{4.3cm} >{\centering}m{3.7em} *{4}{>{\centering\arraybackslash}m{3.7em}@{}}}
			\hline
			\textbf{$\rm {T_{gr}}$, $^{\circ}$C} & 800 & 760 &720 \\
			\hline		
			\textbf{\boldmath{In content $x$ (in planar), \%}} & 12 & 19 & 26 \\
			
			\textbf{Planar} & A & B & C \\
			\textbf{{NRs (460-nm mask)} } & A1& B1 & C1 \\
			\textbf{{NRs (190-nm mask)} } & A2& B2 & C2 \\
			
			\hline
		\end{tabular}
	\end{center}
\end{table}

	NR arrays were examined by scanning electron microscopy (SEM) using a JEOL JSM-7001F microscope. The SEM reveals that typical NRs A1, B1, and C1 have a height and diameter of $\sim$ 1.3 $\mu$m, while NRs A2 -- C2 are about 780 nm in diameter and 2 $\mu$m in height. Representative SEM images of the NRs are shown in Figure~\ref{fig:SEM}. For A1 -- C1 NRs, the  areas of the vertical (10-10) sidewalls and the pyramidal (10-11) facets are comparable, while for NRs A2 -- C2 the pyramidal (10-11) facets are noticeably smaller. The planar (0001) face is practically absent in all studied samples.

	\begin{figure}[b] 
	
	\includegraphics[width=.99\columnwidth]{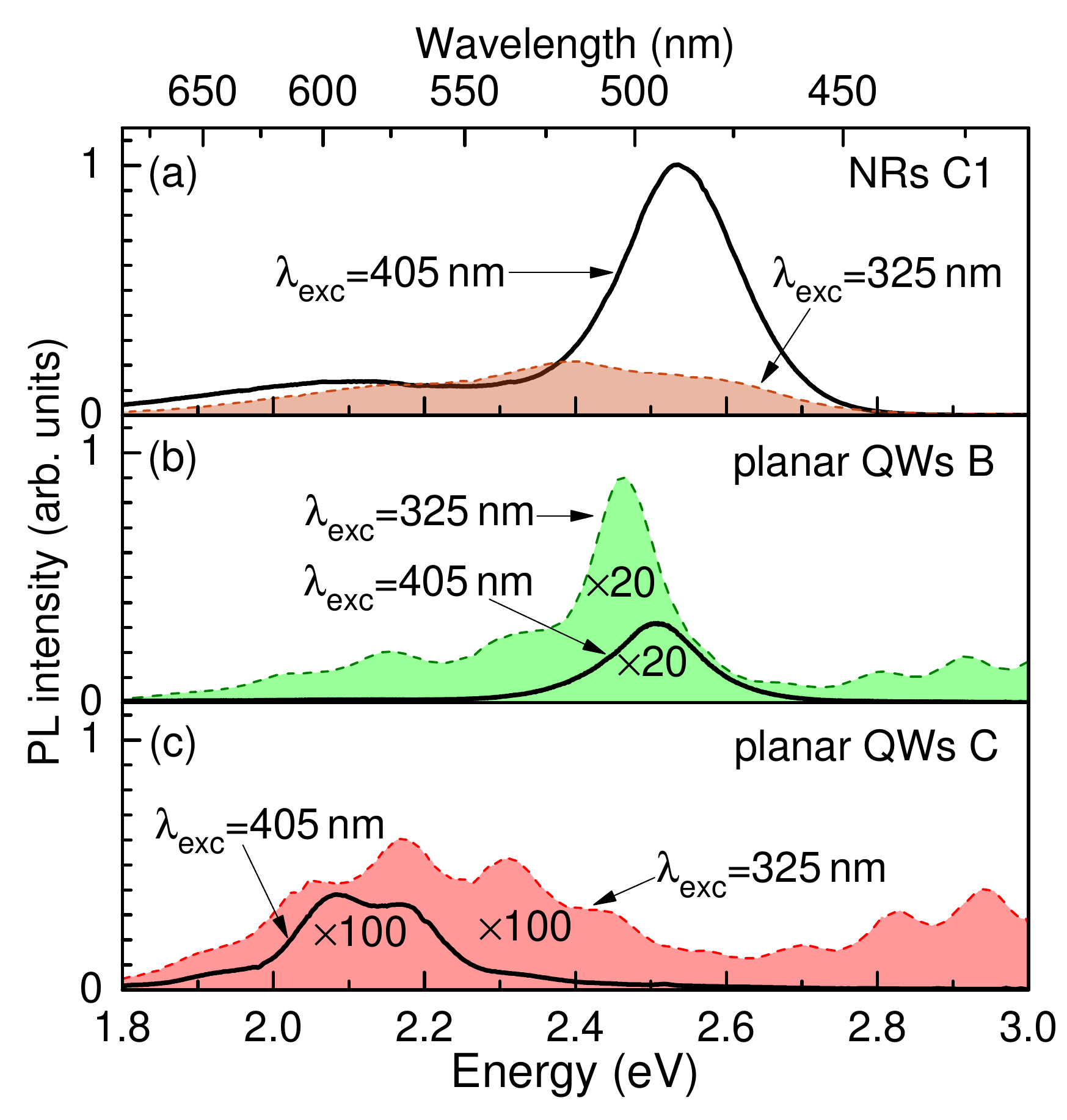}
	\caption{ 
		$\mu$-PL spectra of the NRs C1 (a) with maximal In content and two planar QWs -- B (b) and C (c). Spectra were measured with excitation wavelength $\lambda_{\rm exc}=405$~nm (P$_{\rm exc}=160$~$\mu$W) (solid line) and 325~nm (P$_{\rm exc}=140$~$\mu$W) (dashed lines, filled area). All spectra were measured at RT.
	} 		\label{fig:planar_vs_NR}
	\end{figure}
	
	PL was measured using two setups, either with or without a high spatial resolution. The first setup comprises the LabRAM HR microscope and Horiba Jobin Yvon T64000 triple spectrometer with a liquid nitrogen-cooled charge-coupled detector (CCD). This setup was intended to measure micro-photoluminescence ($\mu$-PL) from single NRs at room temperature (RT). For the excitation of PL we used continues wave (cw) laser lines with wavelengths of 325, 405 and 442 nm. The second setup was used to measure time-integrated PL and TRPL at temperatures from 10 to 300 K. Incident light was focused on the sample by 18 cm focus lens in the geometry close to back-scattering. PL was guided to a spectrometer (Acton2500i) by two achromatic triplet lenses and detected by either CCD or avalanche photomultiplier module (Becker\&Hickl, PMC-150) coupled to a time-correlated single-photon counting system (Becker\&Hickl, SPC130). Long-pass filters were used to suppress scattered laser light in the detection path. The temporal resolution of the TRPL setup was about 140 ps. For optical excitation, we used cw-laser lines with a wavelength of 325 and 377 nm, as well as a line from the pulsed laser with a wavelength of 405~nm and frequency of 30 MHz.

	\section{Results of spectroscopy studies}
	\begin{figure*}[t] 
	\centering
	\includegraphics[width=1.99\columnwidth]{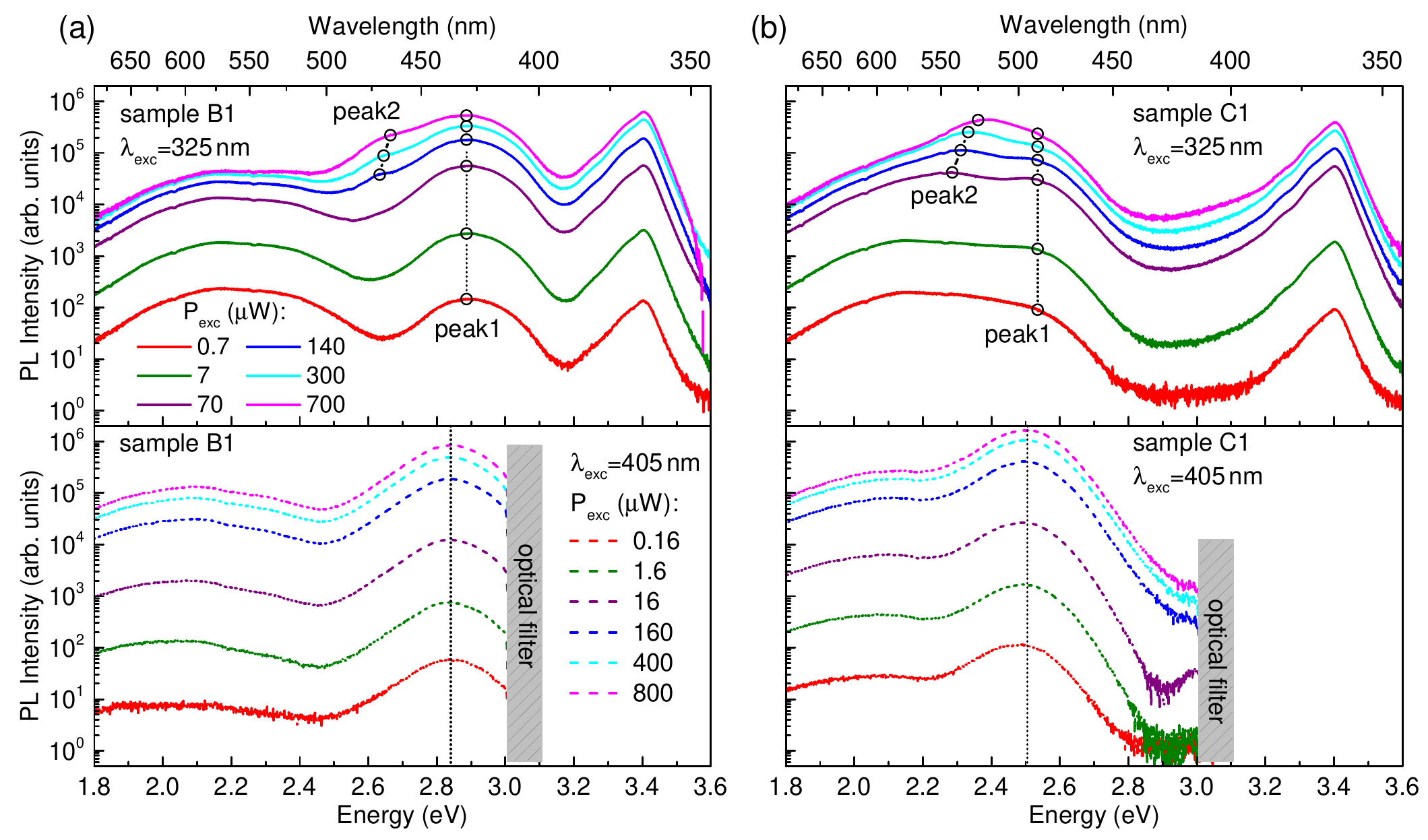}
	\caption{Power-dependent $\mu$-PL spectra measured at RT in the NR samples B1 (a) and C1 (b) excited by cw-laser lines with 325 nm (top panel) and 405 nm (bottom panel) wavelengths.}
	\label{fig:vsP}
\end{figure*}	
	Figure \ref{fig:planar_vs_NR} shows $\mu$-PL spectra of NRs C1 (a) and two planar QWs -- B (b) and C (c) -- for the case of the under-barrier (405 nm) and above-barrier (325 nm) excitations. The PL bands of NRs are blue-shifted and have higher intensity than in the planar QWs of the same In content (note that spectra in (b) and (c) are multiplied by 20 and 100, respectively). Certainly, the NRs provide an apparent gain in PL intensity in the blue-green region, around $\sim$500~nm.

	The PL spectra of NRs also contain a broad yellow-red PL with a peak wavelength of $\sim$580-600~nm, which is usually attributed to the radiation of polar quantum wells. However, its shape  does not depend on the nominal content of In. It was shown in our previous work \cite{2019Evropeitsev}, that the yellow-red emission from NRs with QWs is similar to the spectrum of the yellow band associated with defects, measured in the NRs without QWs.  Such features indicate a significant contribution of defect-related states in GaN to this yellow-red radiation.

	Another PL peculiarity follows from a comparison of the PL spectra from a QW obtained upon above-barrier and under-barrier excitations. The above-barrier excitation is usually more effective due to the enhanced generation of carriers in a thick barrier. However, if their transport towards the QW is hampered,  optical transitions in the barrier will dominate. These two routine situations are realized in the planar sample B with proper transport (Figure~\ref{fig:planar_vs_NR} (b))  and sample C without it (Figure~\ref{fig:planar_vs_NR} (c)). 
	For the C1 NRs, the $\mu$-PL spectrum turns out to be $\sim$5 times less intense for the above-barrier compared with the under-barrier excitation (Figure~\ref{fig:planar_vs_NR} (a)). The shape of NR spectrum has significantly changed -- a  wide yellow-red band dominates and the radiation of the quantum well is completely suppressed. It seems that the emission of defect states in GaN suppresses other recombination channels.  Thus, the question arises whether or not a specific part of the yellow-red PL band \EAQ{is} associated with QWs located in NR.

	\begin{figure*}[t] 
	\centering

		\includegraphics[width=1.99\columnwidth]{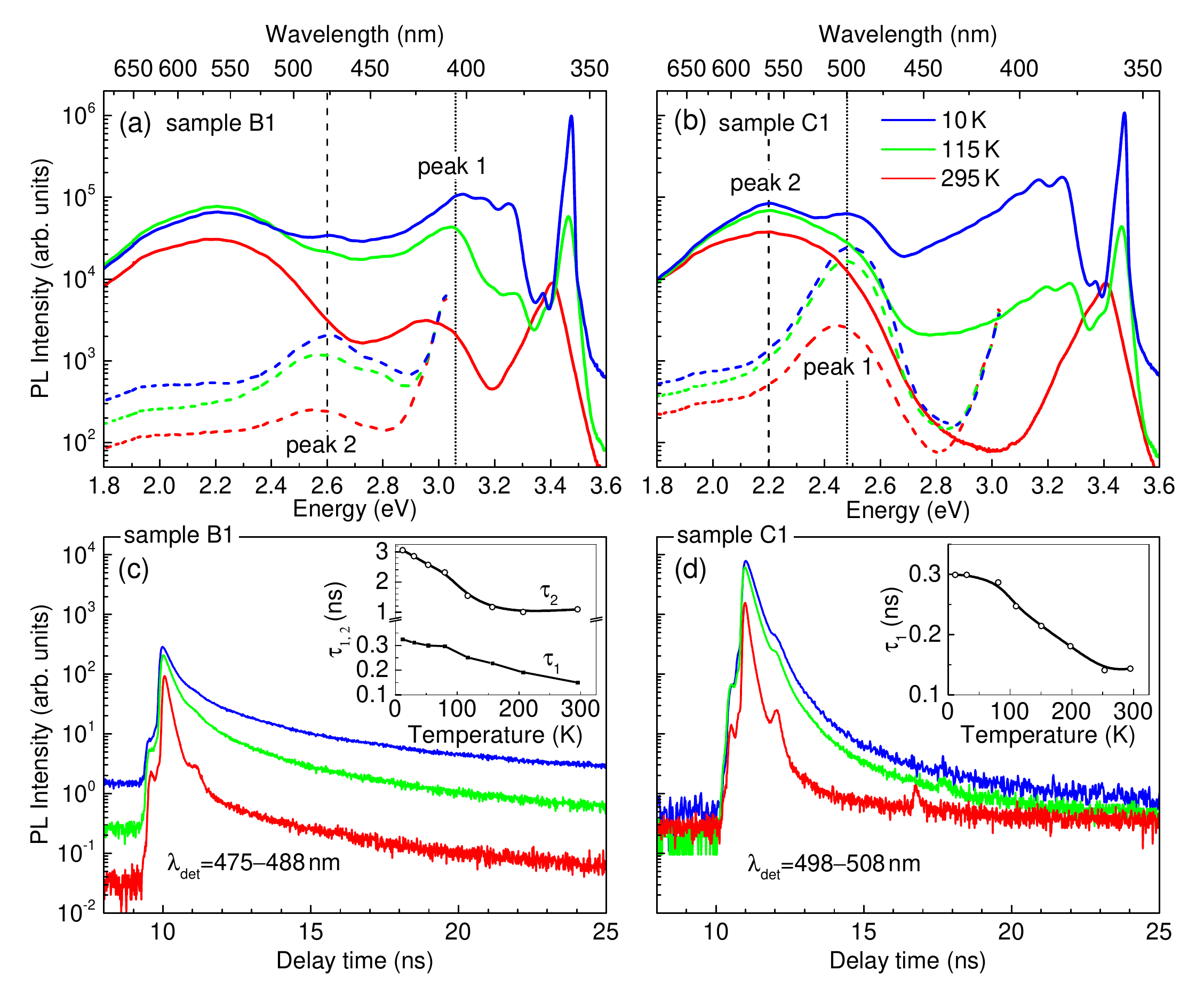}
	\caption{Temperature-dependent PL spectra (a,b) and PL decay curves (c,d) measured in NR arrays B1 (a,c) and C1 (b,d). The PL decay are detected at a wavelength of PL peak excited by a 405~nm laser line with average power of 0.4~mW and repetition rate of 30~MHz.}
	\label{fig:vsT}
	\end{figure*}

	\begin{figure*}[t] 
	\centering
	\includegraphics[width=1.99\columnwidth]{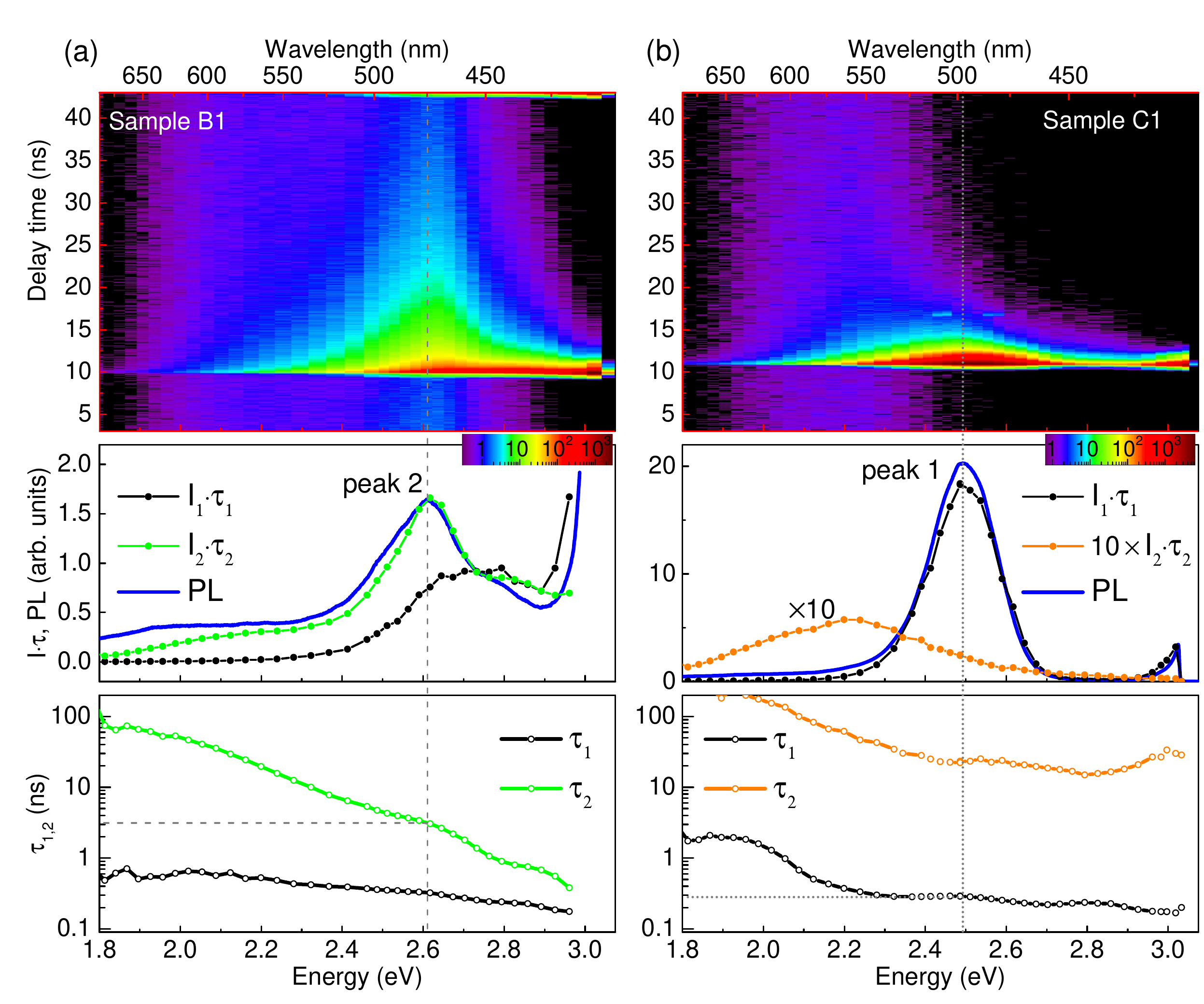}
	\caption{TRPL plots of the  PL intensity  (in logarithm scale marked by color) measured in \EA{NR arrays} B1 (a) and C1 (b) at 15 K (top panels). Bottom panels show the spectral variation of characteristic decay times $\tau_{\rm1}$ and $\tau_{\rm2}$ extracted from the PL decay curves for these samples.   Corresponding spectra of PL (measured by CCD), ${I_1}\cdot{\tau_{\rm1}}$, and ${I_2}\cdot{\tau_{\rm2}}$ are shown in the middle panels.}
	\label{fig:colormap}
	\end{figure*}

	To clarify this problem, we performed power-dependent measurements of $\mu$-PL with excitations above and below the barrier. Typical series of RT $\mu$-PL spectra measured in the single NRs B1 and C1, are shown in Figure \ref{fig:vsP}~(a, b). For $\lambda_{\rm exc}=325$~nm, the spectrally broad yellow-red band dominates at the weak P$_{\rm exc}=0.7$ $\mu$W. However, its intensity saturates with increasing P$_{\rm exc}$. This behavior, typical for defect-related luminescence,
	allows us to conclude that the yellow-red PL upon excitation above the barrier is mainly associated with some defects in GaN.

	In addition, the $\mu$-PL spectra with  $\lambda_{\rm exc}=325$~nm  contain  peak~1, whose wavelength is 390~nm in \EA{NR arrays} A1, 425~nm in B1, and 490~nm in C1. As P$_{\rm exc}$ increases, another peak~2 appears at the wavelengths of $\sim$ 470, 485, and 540~nm, respectively. 	The peak 1 exists even at small powers (0.7~$\mu$W)  and  it does not change its spectral position with increasing P$_{\rm exc}$ to 700~$\mu$W.  On the contrary, peak 2 shifts towards higher energy by 85~meV, 46~meV and 76~meV in \EA{NR arrays} A1, B1, and C1, respectively. With the under-barrier excitation, peak~2 disappears (Figure \ref{fig:vsP}, bottom panel) or decreases significantly  (Figure~\ref{fig:vsT}~(a), bottom panel).  The $\mu$-PL spectra, measured in NRs of a smaller diameter (samples A2, B2, and C2) have similar features. There are some differences, e.g., \EAQ{the decreased} intensity of peak~2. The behavior of peaks 1 and 2 with increasing power allow us to ascribe them to the radiation of the nonpolar and semipolar QWs, respectively. In this case, the observed difference in peak intensities in the samples A1 -- C1 and A2 -- C2 is associated with different areas of (10-11) pyramidal facets.

	The temperature dependencies of macro-PL spectra, measured in \EA{NR arrays} B1 and C1 are shown in Figure~\ref{fig:vsT} (top panels of (a) and (b), respectively). The spectra substantially depend on the $\lambda_{\rm exc}$. For instance, in the sample B1 peak~1 appears at 420~nm upon above-barrier excitation (solid curves), while weak peak~2 appears at 475~nm upon under-barrier excitation (dashed curves) at low temperature (LT). Nevertheless, the PL intensity ratio $I$(295~K)/$I$(10~K) for both  peaks is in the 12-15~\%  range  (P$_{\rm exc}$=15~mW). These temperature dependencies show that the PL intensities of the nonpolar and semipolar QWs dominates at any temperatures with under-barrier excitation. 

	We perform the low-temperature TRPL studies with under-barrier excitation which is more appropriate to reveal the radiation of QWs. We will focus on two samples with highest nominal In content (B1: x = 0.19 and C1: x = 0.26), since red emission requires the high In content in a polar QW.
	The PL intensity dependencies upon the time and energy  are shown in top panels of Figure~\ref{fig:colormap} (a) and (b) for the NR arrays B1 and C1, respectively. They exhibit the presence of both rapidly and slowly decaying components. In order to estimate their particular contributions, the decay curves were fitted by a sum of two exponential functions
	\begin{equation}
	I(t) = I_1 e^{-(t-t_0))/\tau_1} +  I_2 e^{-(t-t_0))/\tau_2},
	\end{equation}
	where $t_0$ is the excitation time ($\sim10$~ns), $\tau_1$ is the fast decay time constant and $\tau_2$ is the slow decay time constant.
	Spectral dependencies of $\tau_i~(i=1, 2)$, obtained for NR arrays B1 and C1, are shown in the bottom panels of Figure~\ref{fig:colormap}. The detection wavelength was set to the maxima of peak~2 (475-488~nm) for B1 and peak~1 (498-508~nm) for C1. The value of  $I_i\cdot\tau_i$ reflects the contribution of $i^{th}$ component to the total PL signal. For sample B1, the spectrum of $I_i\cdot\tau_i$ reveals a dominant contribution of slowly decaying component to peak~2 (see comparison of PL measured by CCD and $I_i\cdot\tau_i$ spectra in the middle panel). On the contrary, for sample C1 only the fast decaying component contributes to peak~1. \EAQ{These are consistent} with the origin of the peak~1 and peak~2 from nonpolar and semipolar QWs, respectively.  The weak slow decaying PL component ($I_2\cdot\tau_2$), red-shifted relatively to the peak~1, may be associated with defect-related transitions.

	The fast decay time $\tau_1$ is equal to 300~ps for both samples B1 and C1 at LT. With temperature rising, $\tau_1$ decreases down to 140-150~ps at RT (see the inserts in Figure~\ref{fig:vsT}~(bottom panels)) that is close to the time resolution of the setup. Since the fast time is dominating in peak 1 at temperatures below 80~K,  we can suggest that it is close to the radiative time $\tau^{rad}$ of these transitions. The slow decay time $\tau_2$ of peak~2 in B1 decreases \EAQ{from 3~ns (at LT) to 1~ns (at 150~K)}. The temperature dependencies of $\tau_1$ and  $\tau_2$ measured in peak~2 indicate that the nonradiative recombination processes are significant even at LT. Therefore we assume that its radiative decay time constant $\tau^{rad}$ may be higher than 3~ns. The obtained values are again consistent with the definition of peaks as emission from nonpolar and semipolar QWs.
	
	To summarize, the spectroscopic studies were intended to differentiate the contributions to the PL spectra from different types of quantum wells - nonpolar, semipolar, and polar - which, as we initially thought, could give a peak~1, a peak~2, and a wide yellow-red band. Indeed, peak~1 does not shift with increasing P$_{exc}$ and has the small value of $\tau^{rad}\approx0.3$~ns. These are important arguments to attribute this peak to recombination in nonpolar QWs. As regards to peak~2, it shifts toward higher energies with increasing P$_{exc}$ and has a radiative decay time of $\tau^{rad}>$3~ns. Therefore, we assume that peak~2 refers to quantum wells with a built-in electric field. Given its energy, these are most likely semipolar quantum wells. The last component, the yellow-red band, quickly saturates with increasing P$_{exc}$ upon above-barrier excitation, being rather weak with under-barrier excitation. These properties are not compatible with QW radiation. In addition, a similar band was observed in NR without QWs. Therefore, we must designate the yellow-red band as predominantly defect-related radiation from GaN.

	\section{Discussion and outlook}
	
	The simultaneous formation of polar, semipolar and nonpolar GaN/InGaN QWs by MOCVD in core-shell nanostructures is a state-of-art method to reach the applications such as while LEDs and multicolor displays. However, the problem of weak red emission is a critical drawback of these structures. We assume that there are two principal factors leading to the weak red emission from polar QWs at the tip of the NR. First and the most obvious is the small area of such QWs comparatively with other QWs.  This drawback was overcome in paper \cite{Robin2018b} inventing wide NRs for red radiation. The second point is the internal electric field of the polar QWs, which decreases a radiative recombination rate. 
	
		\begin{figure*}[t] 
		\centering
		\includegraphics[width=1.99\columnwidth]{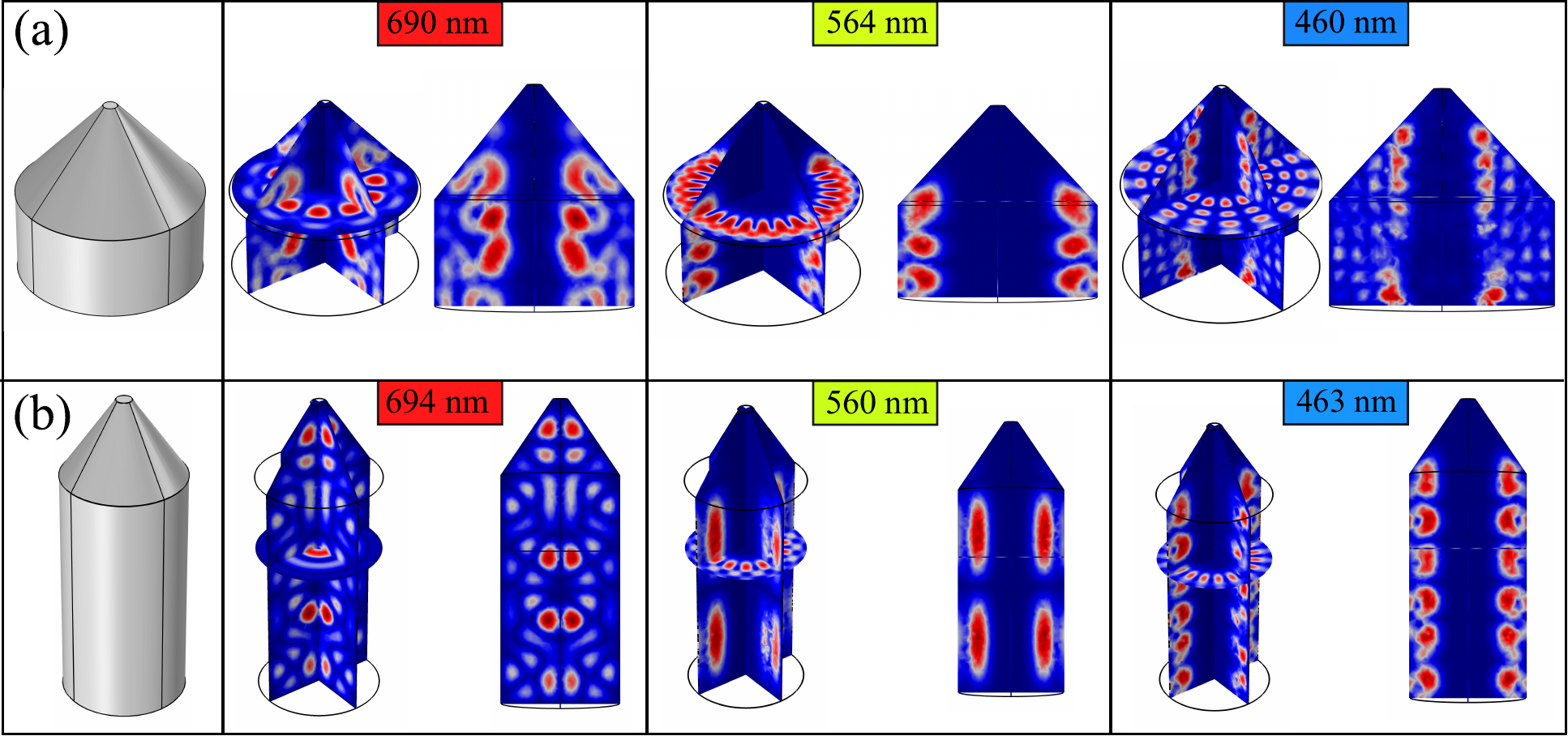} 
		\caption{Images of calculated eigenvalues of the electromagnetic field for two geometries at different wavelengths (R, G, B). Geometry (a) corresponds to wide NRs (type A1) with a cylinder part with radius r$_1 = 650$ nm, height h$_1 = 600$ nm and a conical part with h$_2=650$ nm. Geometry (b) corresponds to narrow NRs (type A2) with a cylinder part with r$_1 = 390$ nm, h$_1 =1500$ nm and a conical part with h$_2=500$ nm. In both cases, upper radius r$_2 = 50$ nm. The red color corresponds to the Purcell factor of optical modes $\geq$100.}
		\label{fig:Modes}
	\end{figure*}

	We consider the NR as a microcavity, where the value of the radiative recombination rate can be increased or decreased due to the Purcell effect under conditions of weak coupling between the confined optical mode and optical transitions inside the cavity. It works by analogy with the Fermi's “golden rule” and depends on the density of photon states and the correspondence of the frequencies of optical and  material resonances \cite{Kleppner1981}. A possible modification of the radiative recombination rate due to the Purcell effect can be expressed as \cite{Gerard1998}
	\begin{equation}
		\frac{\tau^{rad}(\lambda)} {\tau} = \frac{2}{3} F_{\rm P} \frac{|E(r)|^2}{|E_{\rm max}|^2}\frac{\Delta \lambda^2}{\Delta \lambda^2 + 4(\lambda - \lambda_{\rm exc})^2},
	\end{equation}
	where $\tau^{rad}$ and $\tau$ are the intrinsic and changed radiative recombination times, respectively, $\Delta \lambda$ is the spectral width of the resonant mode, $E(r)$ is the electric field distribution, $F_{\rm P}$ is the Purcell factor that depends on the properties of the resonator \cite{Purcell1946}
	\begin{equation}
		F_{\rm P} = 6 \pi (\frac{ c}{n \omega})^3 \frac{{\rm Q}}{V_{\rm eff}}
	\end{equation}
	where ${\rm Q}$ is the quality factor of a resonant mode, $V_{\rm eff}$ is the effective volume of the mode.
	
	The photon density of states can be visualized by calculated electric field distribution in the cavity of definite geometry. As it was previously reported, such a distribution can be complex and inhomogeneous in space inside a monolithic microcavity \cite{Shubina2015,Shubina2016, Kazanov2019}. An increase in the radiative radiation rate ($\tau^{rad}/\tau$)  allows observing the increase in radiation intensity. On the other hand, the intensity can quench when the photon density is low, e.g., when a cavity size is less than a radiation wavelength \cite{Kleppner1981}. 

	We have studied the electric field distribution inside the structures which are similar to the NRs shown in Figure~\ref{fig:SEM}, i.e. the wide A1 and narrow A2. The simulations were implemented via the finite-difference time-domain (FDTD) method using Comsol Multiphysics software. We replace the hexagonal structure by cylindrical one, what does not drastically influence the mode composition \cite{Nobis2004}. The real size cavity was located inside the large cylinder “box” of air with scattering boundaries. This boundary condition allowed us to avoid scattering waves from the “box”, which could create additional interference patterns. The refractive index was chosen to be 	
	\begin{equation}
	n_{\rm mean}(V_1+V_2) = n_1V_1+n_2V_2,
	\end{equation}
	where $\it{n}$$_1$-refractive index of GaN and $\it{n}$$_2$ – refractive index of \EA{In$_{0.15}$Ga$_{0.85}$N}; $V_1$ and $V_2$ are respective volumes. \EA{{However, this value is very close to $\it{n}$$_1$ because of the small volume of QWs.}}

	\EA{{Figure~\ref{fig:Modes} shows “birds-eye” and side views for eigenmodes at different wavelengths of 690 nm, 564 nm and 460 nm. We have chosen these wavelengths for R, G, B because they have the highest Q-factor within these spectral ranges.} }We demonstrate  In the wide NRs of type A1 (a), the electric field distribution at blue and green wavelengths corresponds to the whispering gallery modes (WGMs) of a high order \textit{m}$\geq$20 in the cylinder part, which should increase the radiative recombination rate in nonpolar QWs. The maximum of photon states shifts from the NR boundary to its center with a decrease of the wavelength from 564~nm to  460~nm. The enhancement of PL from semipolar QWs is limited. This wavelength is probably too long for a certain resonator size here; rather, inhibition of spontaneous recombination is possible \cite{Kleppner1981}. Such a situation may be one of the main reasons for the observed inefficient excitation of red radiation from polar QWs in NRs of a conventional design.
	
	In the narrow NRs of type A2, the eigenmode at the wavelength of 460~nm (b) supports the WGMs of a high order \textit{m} $\sim$~20, which should also increase the radiative recombination rate in nonpolar QWs. The 564~nm (green) mode can hardly provide noticeable enhancement for the semipolar QWs. The electromagnetic energy distribution at 690~nm mostly corresponds to the Fabri-Per\'ot like mode, which is concentrated along the center and has of low order $m$. Thus, a noticeable gain in red radiation can only occur if the polar quantum wells are accidentally at the maximum of the Fabri-Per\'ot mode.

	It should be noted that there is a critical radius $r_{cr}$  when the transition occurs from the WGM with $m\neq0$ to the center mode with $m=0$.  The shorter the wavelength, the smaller this radius. For the blue wavelengths, we estimate such a radius as $\simeq$100-150 nm. Below this value the electromagnetic field distribution corresponds to the formation of Fabri-Per\'ot like mode, which are able to provide an increase in spontaneous recombination rate of polar QWs, whilst the blue emission from sidewall nonpolar QWs would be suppressed. Such a critical radius is nothing more than an indicator of the transition from NRs to nanowire. 
	
	Importantly, to improve the situation with red radiation from polar QWs it is possible to form them not only in the tip but also in the core of the NR, as it is shown in Figure~\ref{fig:Design} (b).  The area of such QWs will be obviously increased and their number may be taken as needed. This concept is in line with hybrid nanowires for solar cells, comprising  coaxial and uniaxial InGaN/GaN QWs \cite{Park2018}. We propose the positioning of the uniaxial QW in the maxima of electromagnetic field distribution inside the core to provide  radiation enhancement by the Purcell effect. \EA{{We should note that electromagnetic field distribution will not be affected by the QWs because the volume of each QW is small. The $n_{\rm mean}$ value will change a bit when few QWs are inside the core at the maximum of the Fabry-Per\'ot modes. }} Pumping of such polar QWs will be by UV-blue radiation of sidewall nonpolar QW, similarly to monolithic white LEDs \cite{Damilano2001,Damilano2008}.
	
	Such hybrid design would involve two stages of manufacture. At the first stage, columns with a set of polar QWs designed for the red spectral range can be made from a planar structure using top-down etching technology. In the second stage, a shell with nonpolar QWs emitting in the blue range are grown on the columns. 
	The effect of electromagnetic field distribution should be taken into account for the balance between the enhancement of polar QWs radiation by the Purcell effect and their efficient pumping  by sidewall QWs radiation. With proper designing, such hybrid structure can act as an efficient converter of electrically pumped blue luminescence to the red radiation.

	\section{Conclusions}
	Radiative properties of the core-shell NRs with In$_x$Ga$_{1-x}$N/GaN QWs grown by MOCVD have been investigated by temperature-dependent PL and time-resolved PL, as well as power-dependent $\mu$-PL, performed with under-barrier and above-barrier excitations. For the NRs with nominal In composition of 12, 19, or 26~\%, a pair of PL peaks has been recorded, respectively, in the near-UV, blue and green spectral ranges. We have shown that these peaks arise from nonpolar and semipolar QWs, while the PL from polar QWs is either absent or hidden in the yellow-red emission related mostly to the defect states in GaN. To clarify the red radiation problem, we have considered the NR as an optical cavity, where the Purcell effect can occur. Calculations of the electric field distribution inside the NRs of different sizes have shown no chance for the red radiation enhancement, but rather its quenching  with conventional NR design.  To overcome this problem  we have proposed a hybrid NR design which contains of the polar QWs inside a GaN core, pumped by the radiation of nonpolar QWs. This design is aimed to convert the blue emission from the sidewall nonpolar QWs into the red radiation.  We believe that our results should be taken into account when developing any of a diverse family of NRs for photonic applications.

\bigskip	
	
	$\bf{Acknowledgments}$
	 The work was partly supported by the RFBR grant No. 19-02-00185. The authors thank the Russian Science Foundation (project No. 19-12-00273) for the support of optical modes calculations.

\end{document}